\documentclass[runningheads]{llncs}

\usepackage{eccv}

\usepackage{eccvabbrv}

\usepackage{graphicx}
\usepackage{booktabs}
\usepackage{algorithm,algorithmic}
\usepackage{multirow}
\newcommand{\bacca}[1]{{\color{black}{#1}}}
\newcommand{\brayan}[1]{\textcolor{black}{#1}}
\newcommand{\kebin}[1]{{\color{black}{#1}}}

\usepackage[accsupp]{axessibility}  

\usepackage[pagebackref,breaklinks,colorlinks,citecolor=eccvblue]{hyperref}

\usepackage{orcidlink}

\DeclareMathOperator*{\argmin}{arg\,min}
\newcommand{\blx}{\boldsymbol{x}}
\newcommand{\bly}{\boldsymbol{y}}

\begin{document}

\title{Autoregressive High-Order Finite Difference Modulo Imaging: High-Dynamic Range for Computer Vision Applications}

\titlerunning{AHFD Modulo Imaging}

\author{Brayan Monroy\inst{1}\orcidlink{0000-0002-1087-3062} \and
Kebin Contreras\inst{2} \orcidlink{0009-0000-9122-1893} \and
Jorge Bacca\inst{1}\thanks{This work was supported by VIE-UIS, under project 3968 and 3924.}\orcidlink{0000-0001-5264-7891} }

\authorrunning{B.~Monroy et al.}

\institute{Universidad Industrial de Santander, Bucaramanga, Colombia \and
Universidad del Cauca, Popayán,  Colombia \\ 
\url{https://github.com/bemc22/AHFD}  } 

\maketitle

\begin{abstract}

 High dynamic range (HDR) imaging is vital for capturing the full range of light tones in scenes, essential for computer vision tasks such as autonomous driving. Standard commercial imaging systems face limitations in capacity for well depth, and quantization precision, hindering their HDR capabilities. Modulo imaging, based on unlimited sampling (US) theory, addresses these limitations by using a modulo analog-to-digital approach that resets signals upon saturation, enabling estimation of pixel resets through neighboring pixel intensities. Despite the effectiveness of (US) algorithms in one-dimensional signals, their optimization problem for two-dimensional signals remains unclear. This work formulates the US framework as an autoregressive $\ell_2$ phase unwrapping problem, providing computationally efficient solutions in the discrete cosine domain jointly with a stride removal algorithm also based on spatial differences. By leveraging higher-order finite differences for two-dimensional images, our approach enhances HDR image reconstruction from modulo images, demonstrating its efficacy in improving object detection in autonomous driving scenes without retraining.

\end{abstract}

\section{Introduction}
\label{sec:intro}

High dynamic range (HDR) imaging involves capturing the entire range of light intensity of a scene, preserving \brayan{the details of the image \kebin{ranging from dark to bright areas}} in contrast to under/overexposed images captured by standard digital cameras~\cite{Debevec}.
HDR images play an essential role in the production of high-quality images and videos, as they offer a more authentic and visually appealing representation of the real world. This is particularly important in industries such as photography~\cite{fairchild2007hdr}, cinematography~\cite{tocci2011versatile}, and computer vision~\cite{johnson2003rendering}, where visual fidelity and detail are crucial~\cite{mccann2011art}. Furthermore, HDR imaging of dynamic scenes reduces the dazzle effect caused by sudden changes in lighting conditions, which is essential for vehicle or traffic light recognition in autonomous driving~\cite{wang2016real, wang2018traffic}.

However, standard digital cameras face significant challenges in capturing HDR images due to inherent deficiencies such as limited well capacity and quantization precision~\cite{so2022mantissacam}. Several methods have been explored to capture HDR images using standard digital cameras. For instance, HDR imaging from multiple exposure acquisition involves capturing several images of the same scene at different exposure levels and then combining the pixel information from these images~\cite{10.1145/3596711.3596779, mertens2009exposure, cho2004extending}, which, knowing the sensor response, allows the estimation of an HDR image from these multiple low dynamic range (LDR) images~\cite{10.1145/3596711.3596779}. While multiple exposure capture can produce high-quality HDR images, it is less effective in dynamic scenes with variable lighting, such as high-speed imaging or autonomous driving, where rapidly changing scenes make multiple exposures challenging to use. In response to these limitations, new approaches are emerging that manipulate pixel intensity as a coding strategy enabling HDR image reconstruction from a single-shot LDR image~\cite{hirakawa2011single}. Traditional pixel coding techniques include pixel-wise variable exposure~\cite{nayar2000high} or the incorporation of diffractive optical elements~\cite{sun2020learning, baek2021single, metzler2020deep} 
or recently, modulo sensors~\cite{zhou2020unmodnet,so2022mantissacam}. 

Encoding pixel intensity with smart vision sensors offers a promising strategy to reduce information loss due to saturation, particularly by incorporating the modulo operator in the processing of collected intensity values at the pixel level, also known as modulo imaging~\cite{zhao2015unbounded}. The foundation of modulo imaging involves altering the scene's intensity before sensing it and subsequently reconstructing it using computational methods. Modulo imaging takes advantage of the unlimited sampling framework (USF)~\cite{bhandari2021unlimited}, which states that using modulo analog-to-digital converters (modulo-ADC) that automatically reset or wrap the signal upon reaching a saturation point in continuous time, allows for the estimation of the number of times each pixel is reset using information from adjacent pixel intensities~\cite{bhandari2017unlimited}. USF exploits the idea that spatial differences in a scene are the same as those in modulo measurements, where for the one-dimensional signal, the integration operators are effectively applied~\cite{bhandari2020unlimited}. However, the exact optimization problem addressed within this framework, particularly for high-order finite differences in USF, remains unclear, limiting its generalization or extension to two-dimensional signals.

Due to the correlation between the discontinuities observed in modulo images and the challenges inherent in 2D phase unwrapping, various researchers have adapted phase unwrapping algorithms to address HDR modulo imaging~\cite{bacca2024deep, so2022mantissacam}. However, most of these algorithms predominantly rely on the first spatial finite difference, wasting the advantages offered by higher-order finite differences. In contrast to phase images, the inherent discontinuities in HDR images frequently violated Itoh's condition~\cite{itoh1982analysis}, which limits the image unwrapping using only the first finite-difference.

Consequently, this \kebin{work} proposes to formulate the USF framework as an autoregressive $\ell_2$ phase unwrapping problem and provides computationally efficient analytical solutions in the discrete cosine domain. By adopting this approach, we offer guidelines on how to utilize high-order finite differences for two-dimensional images, thereby improving the reconstruction of HDR images from modulo measurements. Furthermore, from a visual inspection, we notice that the proposed autoregressive method, using multiple differences, provides some stripe artifacts according to the vectorization ordering of the image. As a result, we propose to apply a stride artifact removal algorithm based on the sparsity in the spatial finite difference and employ closed-form solutions based on the cosine transform to remove the stripe artifacts in a non-iterative way. We demonstrate the potential of modulo sensors and the proposed reconstruction algorithm in object detection tasks for autonomous driving scenes, improving detection accuracy under overexposed conditions without retraining.

\section{Background}

The modulo-sensing process consists of analytically resetting the pixel intensity when a saturation threshold $\lambda$ is reached, i.e., the photon count is reset every time the maximum $\lambda$ value is reached (usually $\lambda=2^8-1=255$)~\cite{bhandari2020unlimited}. Assuming a \brayan{vector} form of the signal $\boldsymbol{x}\in \mathbb{R}^n$, the modulo-sensing model for modulo measurements can be mathematically defined as
 \begin{equation}
\boldsymbol{y} =\mathcal{M}_{\lambda}(\boldsymbol{x}) , \label{eqn:model}
\end{equation}
where $\mathcal{M}_{\lambda}(t) =\text{mod}(t,\lambda) $ is known as the modulo operator and  $\boldsymbol{y}$ is modulo measurements. Obtaining $\boldsymbol{x}$ from \eqref{eqn:model} results in an ill-posed problem due to the nonlinearity of the modulo operator. 

\begin{figure*}[!t]
    \centering
    \includegraphics[width=\linewidth]{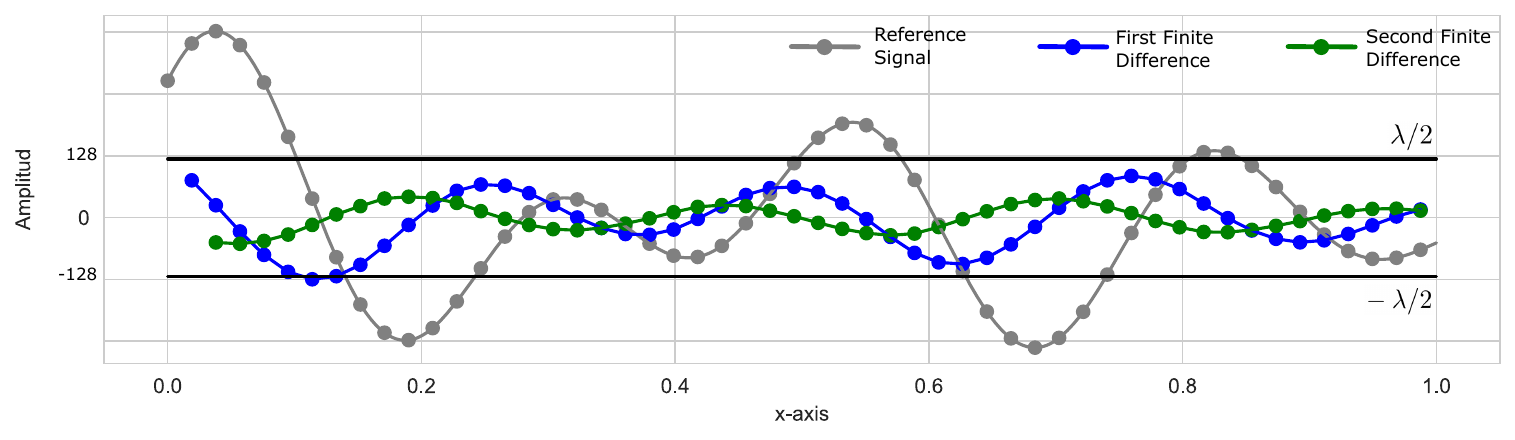}
    \caption{Visual representation of Itoh's condition. For a given band-limited signal, the bounded threshold for the second finite difference its reduced, enabling the unwrapping of modulo samples in contrast to use the first finite difference. Figure inspired from~\cite{bhandari2020unlimited}.}
    \label{fig:itoh}
\end{figure*}

However, following an interesting observation about the finite spatial difference between the signal $\boldsymbol{x}$ and the modulo measurements $\boldsymbol{y}$, it is possible to establish a linear equivalence, guaranteed for some bounded signals, known as Itoh’s condition~\cite{itoh1982analysis}. Specifically, under the condition that $||\Delta \boldsymbol{x}||_{\infty}< \lambda / 2$ the following equation is maintained
\begin{equation}
    \mathcal{M}_{\lambda}(\Delta\boldsymbol{y})=\Delta \boldsymbol{x}.
    \label{eq:difference}
\end{equation}
This formulation of the problem on the finite difference is linear with respect to $\boldsymbol{x}$, and there are several methods available in the field of phase unwrapping to solve it~\cite{yu2019phase, pineda2020spud}. However, in contrast to phase images that are usually smooth, this assumption is difficult to satisfy for a range of images where some spatial differences may be larger than $\lambda/2$, as is the case for discontinuities caused by edges. Nonetheless, a more general assumption based on high-order finite differences can be made~\cite{bhandari2020unlimited}
\begin{equation}
    \mathcal{M}_{\lambda}(\Delta^N\boldsymbol{y})=\Delta^N \boldsymbol{x}.
    \label{eq:assumption}
\end{equation}
As is presented by authors in \cite{bhandari2020unlimited}, and illustrated in Fig.~\ref{fig:itoh} the second spatial difference or even higher order finite differences could reduce the bounded threshold, i.e., more images can satisfy $||\Delta^N \boldsymbol{x} ||_{\infty} < \lambda / 2$ for $N>1$ instead of $N=1$; consequently, recover the unwrapped signal can be made by iterative inversion of the high-order finite differences as is explained in the following section.

\newpage

\subsection{Unlimited Sampling Framework (USF)}

The solution of equation~\eqref{eq:assumption} can be interpreted by analyzing the information loss encountered due to the modulo operator. Specifically, we can decompose the HDR image by the addition of the modulo imaging and the unknown wrapped levels as follows
\begin{equation}
    \blx = \bly + \boldsymbol{k}\lambda, \label{eqn:wrap}
\end{equation}
where $\boldsymbol{k} \in \mathbb{N}^n$ represents an integer-value vector that accounts for the number of times the modulo operator has been applied for each pixel.  $\mathcal{M}_\lambda(\boldsymbol{k}\lambda)=0$.  This interpretation is used in the (USF)~\cite{bhandari2020unlimited} to re-formulate the problem \eqref{eq:problemx} to now recover $\boldsymbol{k}$ instead of $\blx$ since an additional prior about the signal can be incorporated, i.e., $\boldsymbol{k}$ is an integer vector. Therefore, taking into account the assumption in \eqref{eq:assumption} we have that
\begin{equation}
    \Delta^N\boldsymbol{k}=\mathcal{M}_{\lambda}(\Delta^N \bly)-\Delta^N \bly,
    \label{eqn:ho-ass}
\end{equation}
where it can be iteratively solved for each difference, taking into account that $\Delta^N\boldsymbol{k}=\Delta\Delta^{N-1}\boldsymbol{k}$ using the summation operator followed by a projection of $\boldsymbol{k}$ into the integer set, as presented in \cite{bhandari2020unlimited}. \kebin{The USF recovery is summarized in Algorithm ~\ref{alg:usf}, where Lines 1-3 compute the high-order finite differences into the modulo samples $\bar{\boldsymbol{y}}$, estimate the residual differences $\bar{\epsilon}_\lambda$ and initialize the first iteration $\boldsymbol{s}_{(1)}$, respectively.} Lines 4-9 compute the iterative inversion of finite differences and estimated the associated constant bias from each iteration $\eta_{(n)}$. Finally, Line~10 computes the unwrapped signal following Equation~\eqref{eqn:wrap}. Although this methodology has worked for 1D signals, it has limitations when adapting it to the modulo image measurements to obtain the HDR images.

\break

 \begin{algorithm}[!h]  \footnotesize
 \caption{Unlimited Sampling Recovery} \label{alg:usf}
 \begin{algorithmic}[1]
 \renewcommand{\algorithmicrequire}{\textbf{Input:}}
 \renewcommand{\algorithmicensure}{\textbf{Output:}}
 \REQUIRE $\bly = \mathcal{M}_\lambda ( \blx )$, $N \in \mathbb{N} $
 \ENSURE  $\hat{\blx} \approx \blx$
  \STATE Compute $\bar{\bly} = \Delta^N \bly$
  \STATE Compute $\bar{\epsilon}_{\lambda} = \mathcal{M}_\lambda(\bar{\bly}) - \bar{\bly}$
  \STATE Set $\boldsymbol{s}_{(1)} = \bar{\epsilon}_\lambda$
  \FOR {$n = 1$ to $N-1$}
  \STATE $s_{(n+1)} = \mathbf{S}s_{(n)}$
  \STATE Round $s_{(n+1)}$ to  $[ \lambda \mathbb{Z} ]$
   \STATE Compute $\eta_{(n)}$ 
   \STATE $ s_{(n+1)} = s_{(n+1)} +  \lambda \eta_{(n)} $
  \ENDFOR
  \STATE Compute $\hat{\blx} = \bly + \mathbf{S}s_{(N)} $
 \RETURN $\hat{\blx}$ 
 \end{algorithmic} 
 \end{algorithm}

\vspace{-3.5em}

\section{Proposed Method}

\begin{figure*}[!h]
    \centering
    \includegraphics[width=0.99\linewidth]{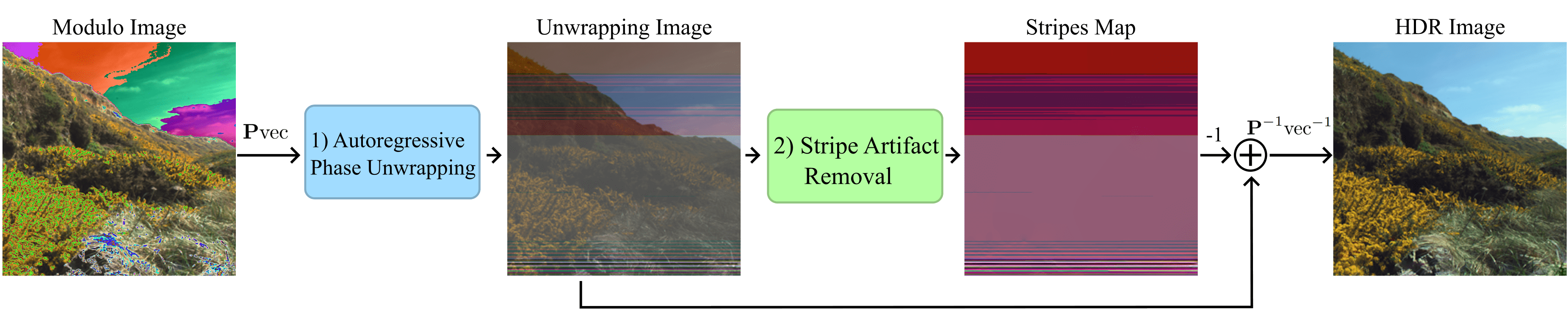} \vspace{-1em}
    \caption{Proposed AHFD method for HDR image restoration from modulo measurements composed of three components: 1) Autoregressive phaseunwrapping algorithm, and 2) Stripe Artifact Removal, 3) The operator $\textbf{P}\texttt{vec}$ to adapt AHFD for matrices. } 
    \label{fig:method} \vspace{-3em}
\end{figure*}

In this section, we present the proposed method called Autoregressive High-Order Finite Difference (AHFD) to obtain HDR images from modulo measurements. The AHFD method comprises three key components: an adaptation of the spatial difference assumption in modulo measurements for matrices (Section \ref{sub:sec:adpatation});  an iterative closed-form solution to handle high-order finite differences based on the  discrete cosine transform (Section \ref{sub:sec:auto-regressive}); and stripe artifact removal based on sparsity in the spatial differences (Section \ref{sub:sec:stripe}). A visual representation of the AHFD method is illustrated in Fig~\ref{fig:method}. \vspace{-1em}

\subsection{Adaptation of USF for Modulo Images}
\label{sub:sec:adpatation}

\begin{figure*}[!h]
    \centering
    \includegraphics[width=0.99\linewidth]{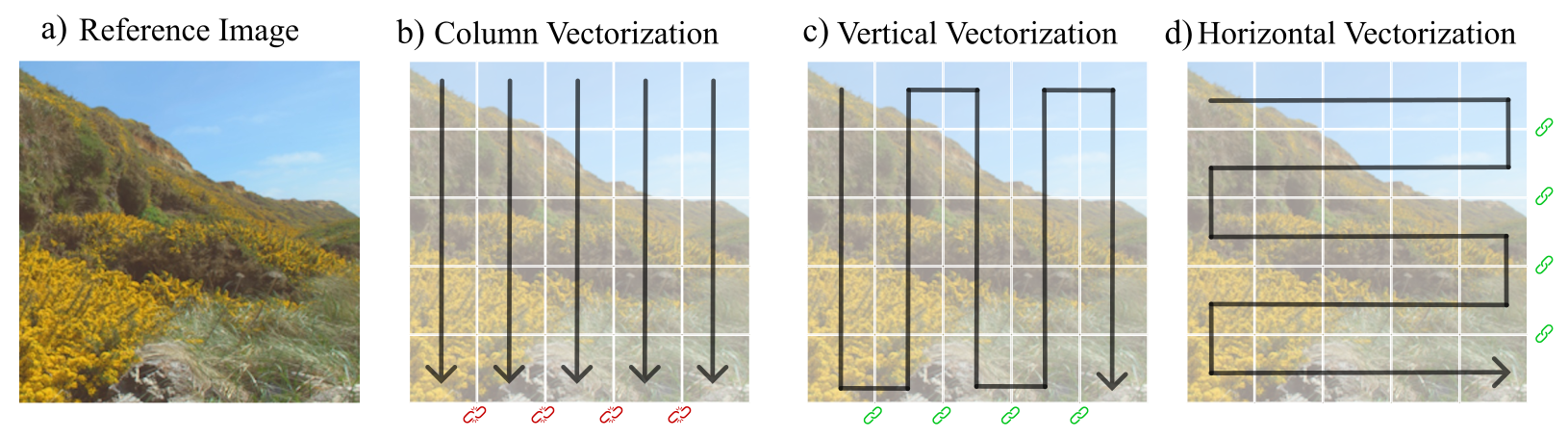} \vspace{-1em}
    \caption{Pixel neighborhood vectorization. a) Reference image, b) column vectorization order using the $\text{vec}(\cdot)$ operator, c-d) proposed pixel neighborhood vectorization $\textbf{P}\text{vec}(\cdot)$ which follows either a vertical or horizontal trajectory. Column vectorization creates artificial discontinuities at the end of each column, whereas our strategy maintains a next-pixel vectorization path. }
    \label{fig:vec} \vspace{-2em}
\end{figure*}

To use USF recovery on 2D modulo measurements, we need to adapt the unwrapping Algorithm ~\ref{alg:usf} of 1d signals for 2d signals. A straightforward alternative could consist of applying the Algorithm~\ref{alg:usf} row-wise or column-wise for each modulo image, however, this approach generates multiple offset artifacts that destroy the spatial image structure~\cite{bacca2024deep}. We propose vectorizing the image and addressing the unwrapping problem for the entire scene to mitigate multiple offsets. However, the traditional vectorization of the matrix $\boldsymbol{x}=\texttt{vec}(\mathbf{X})$, which stacks the columns of the matrix $\mathbf{X}$ on top of one another, increases the likelihood that condition $||\Delta \boldsymbol{x}||_\infty<\lambda/2 $ is not met due to border discontinuities, as illustrated in Fig.~\ref{fig:vec}. Consequently, we proposed an neighborhood vectorization of the form
\begin{equation}
    \boldsymbol{x} =\textbf{P}\texttt{vec}(\textbf{X}),
    \label{eq:vectorization}
\end{equation}
where $\mathbf{P}$ is a permutation matrix that orders the pixels of the image. Equation~\eqref{eq:vectorization} guarantees that the contiguous pixels in the 1d signal are associated with the contiguous pixels in the 2d images. With this vectorization trick, we evaluated the 1d unwrapping recovery as
\begin{equation}
    \tilde{\boldsymbol{x}}\in \argmin_{\boldsymbol{x}} ||
    \Delta^N\boldsymbol{x} - \mathcal{M}_{\lambda}(\Delta^N\boldsymbol{y}) ||_2^2,
    \label{eq:problemx}
\end{equation}
where the HDR image is obtained as $\tilde{\textbf{X}} = \mathbf{P}^{-1}\texttt{vec}^{-1} (\tilde{\boldsymbol{x}})$. Specifially, we proposed two vectorization orderings, following horizontal and vertical vectorization paths as is illustrated in Fig~\ref{fig:vec}. 
\vspace{-1em}

\vspace{-1em}

\subsection{Autoregressive phase unwrapping}
\label{sub:sec:auto-regressive}

\brayan{Similarly to} ~\cite{bhandari2021unlimited}, we proposed to estimate $\boldsymbol{k}$ instead of $\boldsymbol{x}$ based on the assumption presented in Eq~\eqref{eqn:wrap} . Therefore, we proposed the following optimization formulation
\begin{align}
        \argmin_{\boldsymbol{k} \in \mathbb{Z}} ||
\Delta^N\boldsymbol{k}-(\mathcal{M}_{\lambda}(\Delta^N \bly)-\Delta^N \bly) ||_2^2,
\end{align}
using Eq~\eqref{eq:assumption}, i.e., $   \mathcal{M}_{\lambda}(\Delta^N\boldsymbol{y})=\Delta^N \boldsymbol{x}.$
In this sense, we explore the formulation of high-order finite differentiation from USF recovery as an autoregressive phase unwrapping problem. Specifically, based on the fact that $\Delta^N\boldsymbol{k}=\Delta\Delta^{N-1}\boldsymbol{k}=\Delta s^{(1)}$,  
 we can reformulate the problem as a subset of iterative optimization problems where
\begin{equation}
\boldsymbol{s}^{(n+1)}=\argmin_{\Delta^{N-1-n}\boldsymbol{k} \in \mathbb{Z} } \;\; ||
    \Delta(\Delta^{N-1-n}\boldsymbol{k})   -  \boldsymbol{s}^{(n)} ||_2^2 \quad  \text{for}\; n = 0, \cdots, N-1,
    \label{eq:single_problem}
\end{equation}
with $\boldsymbol{s}^{(0)}=\mathcal{M}_{\lambda}(\Delta^N \bly)-\Delta^N \bly$ and $\boldsymbol{s}^{(N)} \approx \boldsymbol{k}$. Interestingly, each sub-problem can also be solved for each component using the DCT transform as the conventional phase unwrapping recovery~\cite{ramirez2024phase}, taking the form
\begin{equation}
s^{(n+1)}_i = \mathcal{D}^{-1}\Big(  \frac{
         \mathcal{D}( \Delta^\top \boldsymbol{s}^{(n)} )_i 
  }{
  2 \cos(\pi i / N ) - 1 
  } \Big).
\end{equation}
Similarly to Algorithm~\ref{alg:usf}, the projection operator into integer set $\mathbb{Z}$ can be employed in each iteration. Our proposed autoregressive phase unwrapping is summarized in Algorithm~\ref{alg:apu}. \vspace{-1em}


\setcounter{algorithm}{0}
 \begin{algorithm}[!h]
 \caption{\textcolor{blue}{Autoregressive Phase Unwrapping}} \label{alg:apu}
 \begin{algorithmic}[1]
 \renewcommand{\algorithmicrequire}{\textbf{Input:}}
 \renewcommand{\algorithmicensure}{\textbf{Output:}}
 \REQUIRE $\bly = \mathcal{M}_\lambda ( \blx )$, $N \in \mathbb{N} $
 \ENSURE  $\hat{\blx} \approx \blx$
  \STATE Compute $\bar{\bly} = \Delta^N \bly$
  \STATE Compute $\bar{\epsilon}_{\lambda} = \mathcal{M}_\lambda(\bar{\bly}) - \bar{\bly}$
  \STATE Set $\boldsymbol{s}^{(0)} = \bar{\epsilon}_\lambda$
  \FOR {$n = 0$ to $N-1$}
  \STATE $s^{(n+1)}_i = \mathcal{D}^{-1}\Big(  \frac{
        \mathcal{D}( \Delta^\top  \boldsymbol{s}^{(n)} )_i 
  }{
  2 \cos(\pi i / N ) - 1 
  } \Big)  $
  \STATE Project $\boldsymbol{s}^{(n+1)}$ into the integer $\mathbb{Z}$
  \ENDFOR
  \STATE Compute $\hat{\blx} = \bly + \boldsymbol{s}^{(N)} \lambda $
 \RETURN $\hat{\blx}$ 
 \end{algorithmic} 
 \end{algorithm}

\vspace{-3em}

\subsection{Stripe Artifact Removal}

\label{sub:sec:stripe}

\begin{figure*}[!t]
    \centering
    \includegraphics[width=\linewidth]{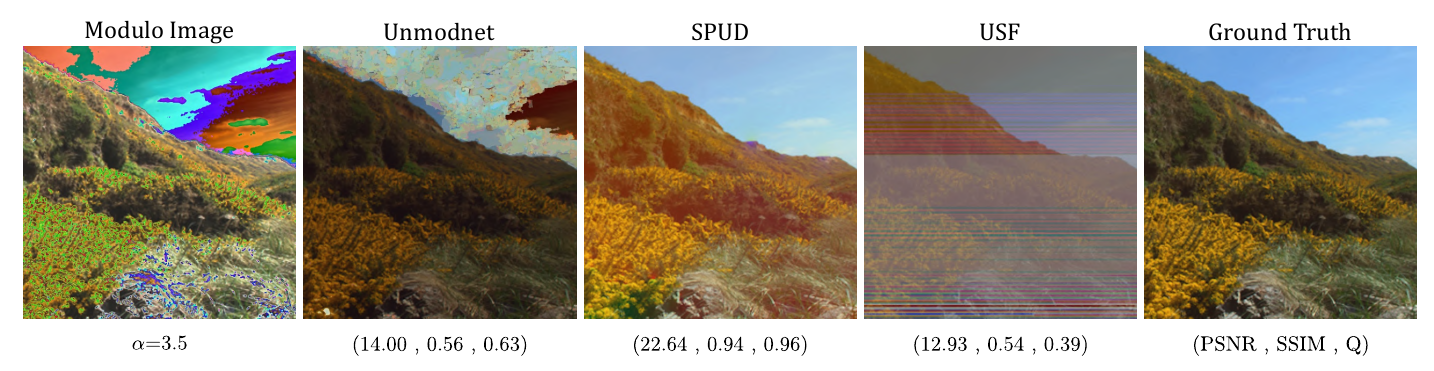} \vspace{-1.5em}
    \includegraphics[width=\linewidth]{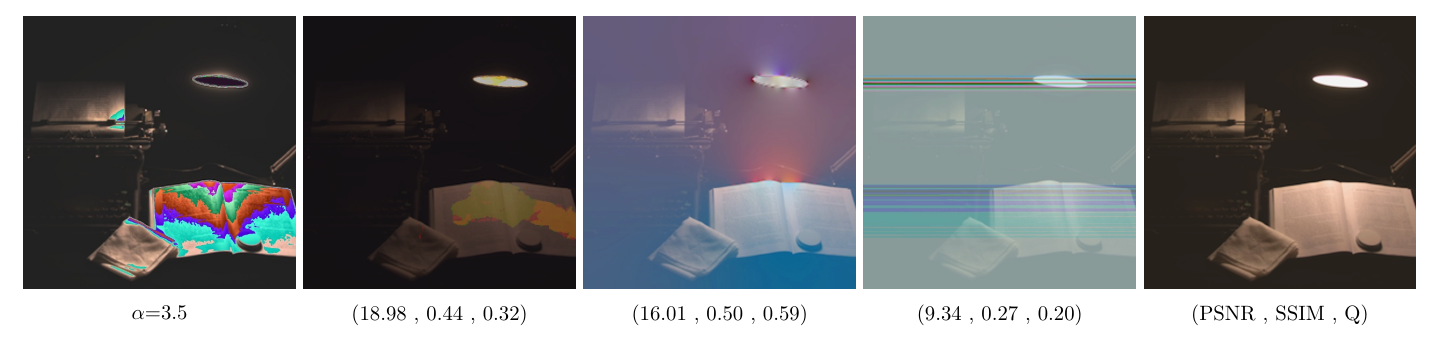} \vspace{-1em}
    \caption{Artifacts from various image unwrapping algorithms. In Undmodnet~\cite{zhou2020unmodnet}, over-estimating unwrapping levels results in "stains" in the images. SPUD~\cite{pineda2020spud} when failure to maintain Itoh's condition causes "light leaks" around synthetic wraps at the image borders. Alternatively, USF~\cite{bhandari2020unlimited} artifacts produce structural lines that overlap with correctly unwrapped images, often referred to as "stripes" in remote sensing images. \vspace{-1em}} 
    \label{fig:artifacts}
\end{figure*}

Our HDR method from modulo measurements in Section~\ref{sub:sec:auto-regressive}, similar to state-of-the-art methods~\cite{ramirez2024phase,bacca2024deep,bhandari2021unlimited}, relies heavily on the assumption of band-limited signals, i.e., $||\Delta \boldsymbol{x}||_\infty<\lambda/2$ is satisfied. When this assumption is violated, ambiguities arise in the high-order finite differences associated with modulo or natural image edges, leading to visual artifacts as shown in Fig.\ref{fig:artifacts}. However, we observe that the artifacts generated using pixel neighborhood vectorization are similar to those found in striping noise\cite{cao2015effective,tsai2008striping}, which are more manageable than artifacts from other phase unwrapping methods, as illustrated in Fig.\ref{fig:artifacts}. Consequently, we propose to address this problem after Algorithm~\ref{alg:apu}. Specifically, the presence of stripe artifacts in the estimated HDR image can be mathematically modeled as\begin{equation}
    \tilde{\boldsymbol{x} } = \boldsymbol{x}  + \boldsymbol{s}.
\end{equation}
Following the idea of spatial difference, we assume that the stripes are sparse and exhibit high frequency in the spatial difference dimensions. Consequently, we propose estimating the stripes by solving the following optimization problem:
\begin{equation}
    \tilde{ \boldsymbol{s} } \in \underset{\boldsymbol{s} \in \mathbb{R}}{\text{arg min}} \quad \Vert 
 \Delta_{xy} \tilde{\boldsymbol{x}}  - \Delta_{xy} \left(\boldsymbol{x+s}\right) \Vert_2^2 + \gamma \Vert \Delta_{x,y} \boldsymbol{s} \Vert_0,
 \label{eq:idea_stride}
    \end{equation}
where $\Delta_{xy} = [ \Delta_x^\top, \Delta_y^\top]^\top $. This  problem represents a conventional $\ell_2 - \ell_0$ optimization, wherein compressive sensing techniques advocate substituting for a $\ell_2 - \ell_1$ formulation and resolving the equations through iterative methods~\cite{bioucas2007new,monroy2022jr2net}. However, following the idea of a non-iterative process~\cite{bacca2019noniterative}, we propose to assume that $\Delta_{xy}\boldsymbol{s}=\Delta_{xy}(\boldsymbol{\tilde{x}}-\boldsymbol{x})=\mathcal{H}_\gamma( \tilde{\blx})$, where $\mathcal{H}_\gamma $ is the hard thresholding operator defined element-wise as $\mathcal{H}_\gamma( \blx)_i = \blx_i \cdot \textbf{1}_{|\blx| \leq \gamma }$ for an input vector $\blx$. Consequently, Eq.~\eqref{eq:idea_stride} can now be expressed as.

\begin{equation}
    \tilde{ \boldsymbol{s} } \in \underset{\boldsymbol{s} \in \mathbb{R}}{\text{arg min}} \quad \Vert 
 \Delta_{xy} \mathcal{H}_\gamma( \tilde{\blx})  - \Delta_{xy} \boldsymbol{s} \Vert_2^2.
 \label{eq:proposed}
    \end{equation}

Interestingly, this problem results in a 2D generalization of a single problem such as in Eq~\eqref{eq:single_problem} and can be solved using 2D DCT~\cite{ramirez2024phase} instead of 1D DCT as summarized in Algorithm~\ref{alg:stripes}. With the inclusion of Algorithm~\ref{alg:stripes}, our proposed method significantly overcomes the performance in cases where modulo measurements present high ambiguity between modulo edges and image edges.

Finally, the proposed modulo recovery algorithm that combines a vectorization operator $\textbf{P}$, the auto-regessive phase unwrapping algorithm, and the strip artifact removal algorithm are summarized in Algorithm~\ref{alg:ours}. The proposed method computes ordered vectorization $\boldsymbol{y}$ for a given modulo image $\textbf{Y}$ (Line 1) and subsequently applies Algorithms~\ref{alg:apu}-\ref{alg:stripes} (Lines 2-3); finally, the estimated vector is reshaped to match the image dimensions. For color images, Algorithm~\ref{alg:ours} is broadcast along the channels dimensions. In terms of time complexity, the main computational bottleneck manifests itself during iterative integration within the Discrete Cosine Transform (DCT) domain for the vectorized modulo image, as detailed in Algorithm~\ref{alg:apu}. This process exhibits a computational complexity of $\mathcal{O}(Kn\log(n))$, where $n$ denotes the total pixel count of the modulo image and $N$ represents the number of finite high-order differences.

 \begin{algorithm}[!t]
 \caption{\textcolor{ForestGreen}{Stripe Artifact Removal Algorithm}} \label{alg:stripes}
 \begin{algorithmic}[1]
 \renewcommand{\algorithmicrequire}{\textbf{Input:}}
 \renewcommand{\algorithmicensure}{\textbf{Output:}}
 \REQUIRE $\tilde{\blx}, \gamma$
 \ENSURE  $\hat{\blx} \approx \blx $
  \STATE  $\boldsymbol{\rho} = \texttt{DTC-2D}(\Delta_{xy}^\top \mathcal{H}_\gamma ( \Delta_{xy}\tilde{\blx} ) )$
  \STATE  $\tilde{\boldsymbol{n}}_{mn+n} =  \texttt{DTC-2D}^{-1}( \boldsymbol{\rho} \cdot [
   2 \cos ( \frac{\pi m}{M}) + 2 \cos ( \frac{\pi n}{N} ) - 4  ] ^{-1} ) $
  \STATE  $\hat{\blx} = \tilde{\blx} - \tilde{\boldsymbol{n}}$
 \RETURN $\hat{\blx}$ 
 \end{algorithmic} 
 \end{algorithm}

 \begin{algorithm}[!t] 
 \caption{Proposed Modulo Recovery Algorithm} \label{alg:ours}
 \begin{algorithmic}[1]
 \renewcommand{\algorithmicrequire}{\textbf{Input:}}
 \renewcommand{\algorithmicensure}{\textbf{Output:}}
 \REQUIRE $\textbf{Y} = \mathcal{M}_\lambda ( \textbf{X} )$, $N \in \mathbb{N}, \gamma $
 \ENSURE  $\tilde{\boldsymbol{X}} \approx \textbf{X} $
  \STATE  \text{Computes} $\boldsymbol{y} =  \textbf{P}\texttt{vec}(\textbf{Y}) $
  \STATE  \text{Computes} $\hat{\blx}$ from Algorithm~\ref{alg:apu}. \quad \textcolor{gray}{\#Autoregressive Phase Unwrapping }
  \STATE  \text{Computes} $\tilde{\blx}$ from Algorithm~\ref{alg:stripes}. \quad \textcolor{gray}{\#Stripes Artifact Removal}
   \STATE  \text{Computes} $\tilde{\textbf{X}} = \texttt{vec}^{-1}(\textbf{P}^{-1} \tilde{\blx})$
 \RETURN $\tilde{\textbf{X}}$ 
 \end{algorithmic} 
 \label{alg:recon} 
 \end{algorithm}

\section{Simulations and Results} \vspace{-0.5em}

This section presents a detailed analysis of the performance and efficacy of the proposed methodology through various experiments on HDR image restoration \bacca{and object detection in autonomous driving scenes}. The following subsections will detail the experimental settings, the evaluation metrics used for the assessment, and a comprehensive overview of the results obtained \bacca{for both tasks}.

For the validation of the proposed \bacca{recovery} method in HDR image reconstruction from modulation \bacca{measurements}, we conducted a comparison with state-of-the-art optimization and deep learning-based unwrapping algorithms. Specifically, for optimization algorithms, we selected the simultaneous phase unwrapping and denoising algorithm (SPUD)~\cite{pineda2020spud} and the Plug-and-Play Unwrapping Algorithm (PnP-UA)~\cite{bacca2024deep}; they proposed non-iterative and deep prior iterative solutions for the phase unwrapping problem \bacca{based on the} first spatial finite difference, respectively. In the case of deep learning-based \bacca{recovery network}, we select Unmodnet~\cite{zhou2020unmodnet}, which employs an iterative estimation of the binary mask to estimate the wrapping levels for modulo \bacca{measurements} and includes Laplacian in the modulo image as guided information. 

\bacca{On the other hand, we employed YOLOv10x~\cite{wang2024yolov10} directly to the measurements to evaluate the object detection task. Specifically, we simulate acquisition of saturated images with a CCD sensor, modulo measurements, and finally, HDR image reconstruction using the proposed method. For this experiment, the network weights are the same for all scenarios and were obtained from ~\cite{wang2024yolov10}.}

\vspace{-1em}

\subsection{Validation on HDR Image Restoration.}

\textbf{Dataset.} We use the Korshunov dataset as a benchmark for HDR image restoration. This dataset includes 20 HDR images with resolutions varying from full HD ($1920 \times 1080$) to greater than 4K ($6032 \times 4018$) and 12 bits per color channel \cite{korshunov2015subjective}. The images encompass various scenes, dynamic ranges, and acquisition techniques, featuring architecture, landscapes, portraits, and CGI-generated visuals. All images were resized to a resolution of $1024\times1024$. \bacca{It is important to highlight that the proposed method is non-data dependent. Therefore, this dataset is only used for testing purposes. }

\textbf{Metrics.} To evaluate the quality of the unwrapped restoration, we utilized three specific metrics: Q-index, peak signal-to-noise ratio (PSNR), and structural similarity index (SSIM). The Q-index is frequently used in phase unwrapping methods, whereas PSNR and SSIM are common image quality metrics. All metrics were applied within the HDR range as per ~\cite{pineda2020spud}. The Reinhard tone mapping function is used for image visualization~\cite{reinhard2023photographic}.

\setlength{\tabcolsep}{5pt}

\begin{table}[!b]
\caption{Quantitative comparison results using different saturation factors for state-of-the-art recovery methods from modulo measurements. Our-h and Ours-v stands for our proposed recovery algorithm with horizontal and vertical vectorization, respectively.}
\label{table:saturations} \vspace{-1em}
\centering  
    \resizebox{\textwidth}{!}{  
    \begin{tabular}{ll|rrrrrr}
        \toprule
         &  &  &  & Saturation & Factor & &  \\ \hline
        Metric & Method &  1.825 & 2.15 & 2.475 & 2.8 & 3.0 & 3.2 \\
        \midrule
        \multirow[t]{6}{*}{PSNR $(\uparrow)$ } & Unmodnet~\cite{zhou2020unmodnet} & 22.77 & 21.93 & 19.63 & 19.26 & 18.78 & 19.29 \\
         & SPUD~\cite{pineda2020spud} & 57.63 & 57.48 & 47.24 & 36.32 & 32.55 & 29.07 \\
         & PnP-UA~\cite{bacca2024deep} &  61.26 & \textbf{60.85} & \textbf{49.78} & 37.01 & 33.11 & 29.53 \\
         & Ours-h & 60.37 & 59.80 & 44.58 & \textbf{43.19} & 36.26 & 32.81 \\
         & Ours-v & \textbf{61.86} & 59.24 & 40.03 & 43.61 & \textbf{37.96} & \textbf{34.05} \\
        \cline{1-8}
        \multirow[t]{6}{*}{SSIM $(\uparrow)$  } & Unmodnet~\cite{zhou2020unmodnet} & 0.6844 & 0.6907 & 0.5602 & 0.5657 & 0.5073 & 0.5897 \\
         & SPUD~\cite{pineda2020spud} & 0.9993 & 0.9993 & 0.9861 & 0.9299 & 0.8826 & 0.8410 \\
         & PnP-UA &  0.9995 & \textbf{0.9996} &\textbf{0.9869}  & 0.9310 & 0.8837 & 0.8422 \\
         & Ours-h & 0.9994 & 0.9994 & 0.9735 &\textbf{ 0.9913} & 0.9607 & \textbf{0.9364} \\
         & Ours-v & \textbf{0.9996} & 0.9992 & 0.8697 & 0.9742 & \textbf{0.9333} & 0.8791 \\
        \cline{1-8}
        \multirow[t]{6}{*}{Q-score $(\uparrow)$ } & Unmodnet~\cite{zhou2020unmodnet} & 0.6194 & 0.6420 & 0.4853 & 0.4852 & 0.4189 & 0.5010 \\
         & SPUD~\cite{pineda2020spud} & 0.9999 & 0.9999 & 0.9904 & 0.9400 & 0.9027 & 0.8527 \\
         & PnP-UA~\cite{bacca2024deep} & 0.9999 & 0.9999 & \textbf{0.9906} & 0.9404 & 0.9031 & 0.8530 \\
         & Ours-h & 0.9999 & \textbf{0.9999} & 0.9777 & \textbf{0.9957} & \textbf{0.9412} & \textbf{0.9418} \\
         & Ours-v & \textbf{1.0000} & \textbf{0.9999} & 0.7547 & 0.9783 & 0.8933 & 0.8265 \\
        \cline{1-8}
        \bottomrule
\end{tabular}
    }
    
    \label{tab:results} \vspace{-1em}
\end{table}

\textbf{Simulations.} The ability of the proposed reconstruction method was evaluated by comparing it against SPUD~\cite{pineda2020spud}, Unmodnet\cite{zhou2020unmodnet}, and PnP-UA~\cite{bacca2024deep}.  For this experiment,  different illumination conditions were evaluated by changing the saturation faction under the synthetic testing dataset. The generation of this dataset is represented as\begin{equation}
\textbf{Y} = \mathcal{M}_\lambda(\alpha \textbf{X}),
\label{equ:satu}
\end{equation}where \( \textbf{X} \) is a normalized HDR image within the range [0, 1] with three channels (RGB) and $1024 \times 1024$ spatial resolution, \( \alpha > 1\) is the saturation factor, and \( \lambda = 1 \). Our evaluations cover a variety of saturation factors, specifically \( \alpha \in \{1.825, 2.15, 2.475, 2.8, 3.0, 3.2\} \).

\begin{figure*}[!h]
    \centering
    \includegraphics[width=\linewidth]{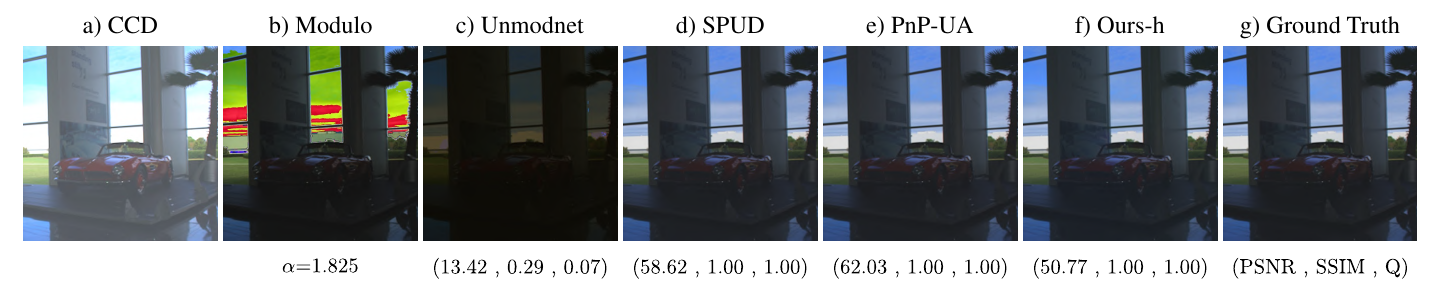}
    \includegraphics[width=\linewidth]{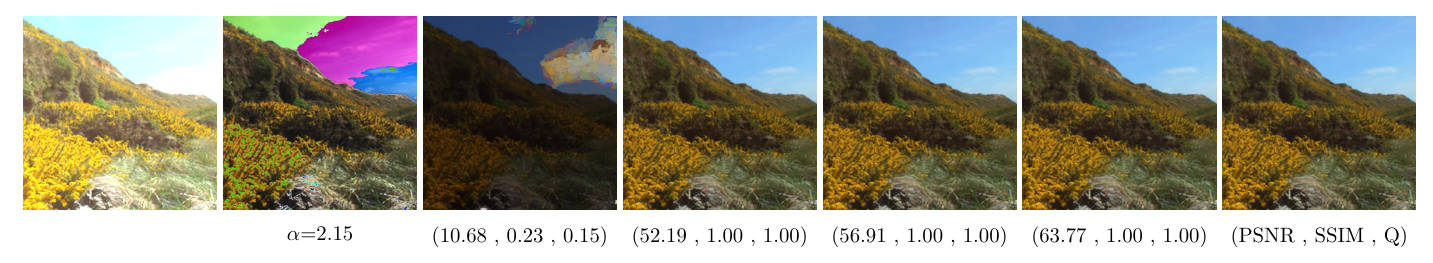}
    \includegraphics[width=\linewidth]{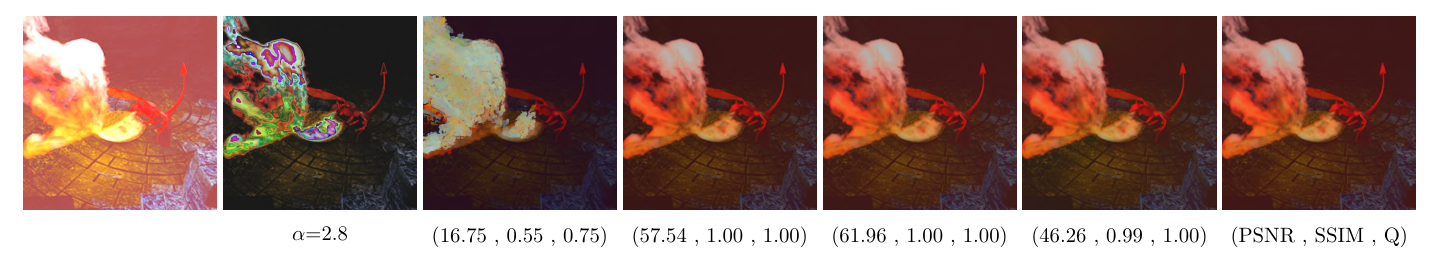}
    \includegraphics[width=\linewidth]{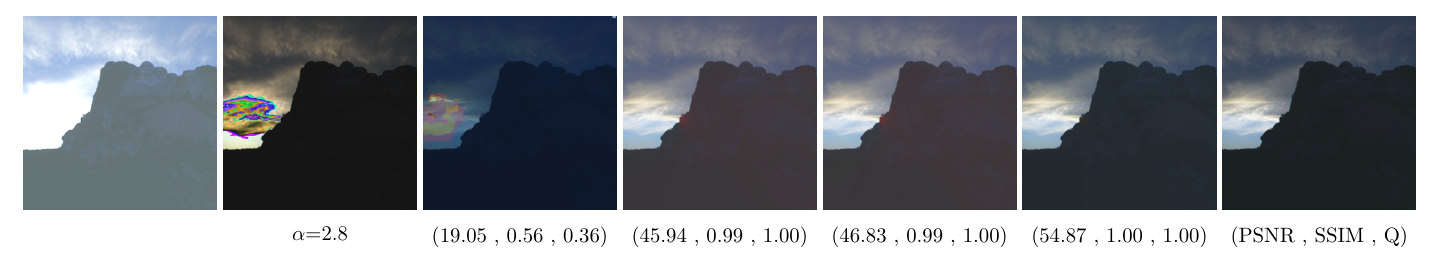}
    \includegraphics[width=\linewidth]{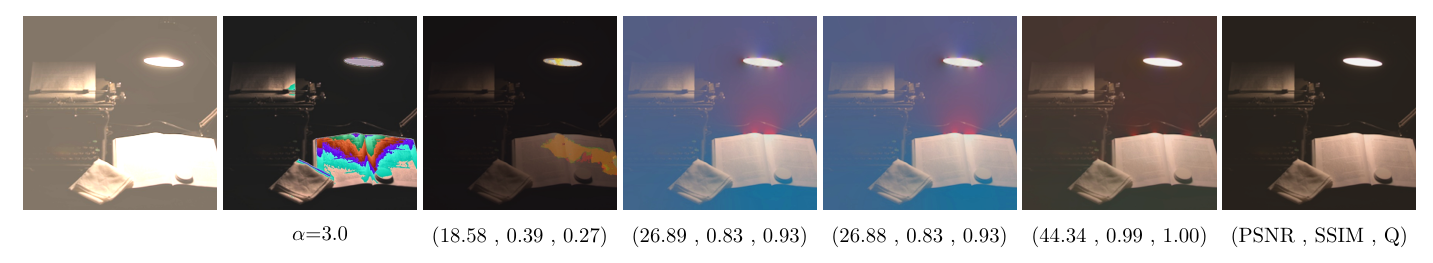}
    \includegraphics[width=\linewidth]{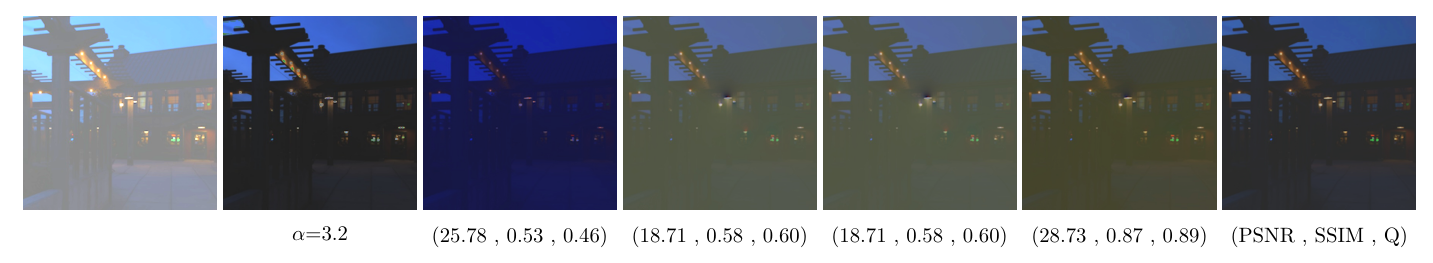}
    \includegraphics[width=\linewidth]{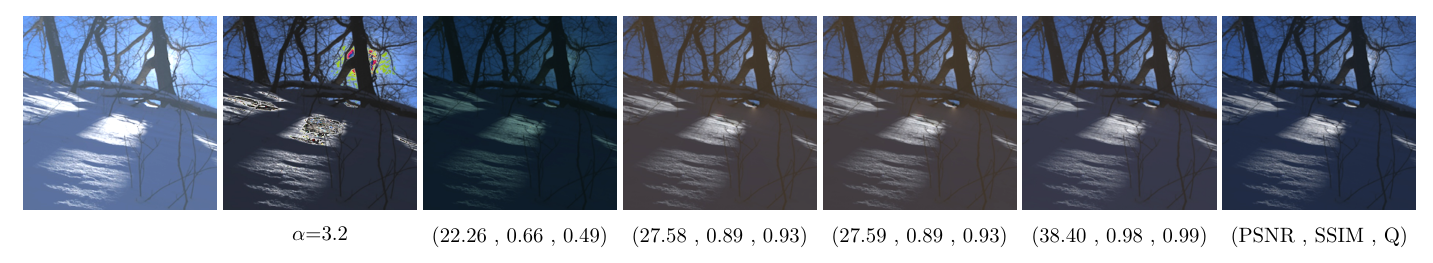}
    \caption{Comparison with state-of-the-art recovery methods from the modulo images. (a-b) correspond to saturated and modulo measurements under different intensity levels, respectively. (c-f) Corresponding to different recovery methods and the proposed Ours-h method, finally, (g) corresponds to the ground truth image.}
    \label{fig:visual}
\end{figure*}

Quantitative results are summarized in Table~\ref{table:saturations}. Specifically, our proposed method delivers performance comparable to SPUD and PnP-UA for $\alpha$ values between $[2.15, 2.475]$ and achieves better results for $\alpha \geq 2.8$. For deep learning methods such as Unmodnet, we observed a poor generalization across all illumination conditions. These quantitative results are supported by Figure~\ref{fig:visual}, which shows that our method surpasses state-of-the-art methods under almost all illumination conditions. At high saturation, it effectively mitigates unwrapping artifacts in areas with high wrapping concentration, such as near light sources like lamps or the Sun. Despite this, there is still room for improvement, as seen in the case of $\alpha=3.0$ in Figure~\ref{fig:visual}, where unwrapped artifacts are still present in our proposed method.

Finally, we present a computational resource consumption analysis in Table~\ref{tab:timememory}. CNN-based methods like Undmodnet and PnP-UA have high memory usage and long running times due to deep neural model parameters, with Undmodnet being the slowest. In contrast, SPUD is the fastest and most memory-efficient due to its non-iterative and low hyperparameter design. Our method is the second fastest and most memory-efficient, with significant performance improvement. This analysis underscores the limitations of CNN-based HDR image restoration on edge devices, whereas our algorithm offers a computationally efficient alternative with competitive HDR restoration results. \vspace{-1.5em}

\begin{table}[!h]
    \centering \caption{Time and Memory consumption between proposed and state-of-the-art HDR modulo imaging methods. All method were tested on a RTX 3090 and R5 5600X CPU. } 
            \begin{tabular}{l||l|l|l|l}
    \hline
    Method        & Undmodnet &  PnP-UA & SPUD & Ours-h/v \\ \hline  \hline
    Running Time {[}ms{]} & 5327        &  456  & 5.51     & 8.34     \\
    VRAM {[}MiB{]} & 2697        & 6717 & 70     & 151       \\ \hline
    \end{tabular}
    \label{tab:timememory} \vspace{-3.5em}
\end{table}

\subsection{Validation on Object Detection for Autonomous Driving.}

\textbf{Dataset.} We select the KITTI database, a reference in \bacca{object detection task} due to its focus on autonomous driving. This database contains 7,481 images with dimensions of $1242\times375$ and covers urban scenes with a variety of objects such as vehicles, pedestrians, cyclists, cars, cyclists, miscellaneous objects, pedestrians, people sitting, trams, trucks, and vans, providing a real challenge for vision-based object detection systems \cite{Geiger2012CVPR}.

\textbf{Model.} The efficacy of object detection was evaluated using the YOLOv10x model, a neural network architecture designed for real-time object detection with 29.5 million parameters\cite{wang2024yolov10}. This model consists in a convolutional neural network backbone with multiple detection layers that utilize anchor boxes to predict bounding boxes for objects within an image. The architecture of YOLOv10x incorporates a consistent dual assignment design that eliminates the need for non-maximum suppression during inference, improving efficiency. Additionally, it employs a lightweight classification head, spatial-channel decoupled downsampling, and a rank-guided block design to decrease computational redundancy and enhance efficiency. It also uses large kernel convolution and a partial self-attention module to improve feature extraction capability and localization accuracy.

\textbf{Metrics.} The performance of the \bacca{object detection task} is evaluated using three specific metrics: Accuracy, Intersection over Union (IoU), and the F1 score. These metrics serve as standards in the evaluation of object detection and enable an objective comparison of the performance of the YOLOv10x model under different image manipulation conditions.

\textbf{Simulations.} The results were obtained by varying the saturation factor $\alpha$ with three specific values: $\alpha = \{ 1.5, 2, 3 \} $, in accordance with Equation \ref{equ:satu}, which adjusted the saturation intensity of the images. The images from the KITTI dataset were processed by simulating four scenarios and then used as input in YOLOv10x to obtain the boundary detection boxes. Specifically, we simulated the acquisition of i) \textbf{CCD}, which refers to the saturated image based on the specific saturation factor, ii) \textbf{Modulo}, which refers to the modulo \bacca{measurements} obtained using Eq.~\ref{equ:satu}, iii) \textbf{Recovery}, which refers to the reconstructed HDR from the modulo \bacca{measurements using Algorithm~\ref{alg:ours}}, and iv) \textbf{Ideal HDR}, which refers to the original image without modification, assuming a perfect acquisition of the HDR image.

\begin{table}[!b]
\centering
\caption{Object Detection Evaluation with Different Saturation Levels} \vspace{-1em}
\label{tab:metodos_recuperacion}
\resizebox{\textwidth}{!}{  
\begin{tabular}{@{}lccccccccc@{}}
\toprule
\multirow{4}{*}{\textbf{Method}} & \multicolumn{9}{c}{\textbf{Metrics}} \\ 
\cmidrule{2-10}
& \multicolumn{3}{c}{\textbf{$\alpha = 1.5$}} & \multicolumn{3}{c}{\textbf{$\alpha = 2$}} & \multicolumn{3}{c}{\textbf{$\alpha = 3$}} \\ 
\cmidrule(lr){2-4} \cmidrule(lr){5-7} \cmidrule(lr){8-10}
& IOU $(\uparrow)$ & F1 $(\uparrow)$  & ACC $(\uparrow)$ & IOU $(\uparrow)$ & F1 $(\uparrow)$ & ACC $(\uparrow)$ & IOU $(\uparrow)$ & F1 $(\uparrow)$ & ACC $(\uparrow)$ \\  \midrule
CCD & 64.69 & 47.05 & 36.70 &     61.56 & 45.67 & 34.45 &          59.60 & 37.98 & 28.29 \\
Modulo  & 74.59 & 61.15 & 50.41 & 73.04 & 58.86 & 47.93 & 70.40 & 55.42 & 44.70 \\
Recovery  & \textbf{74.61} & \textbf{61.25} & \textbf{50.64} & \textbf{74.44} & \textbf{60.95} & \textbf{50.28} & \textbf{71.06} & \textbf{55.86} & \textbf{44.92} \\
\midrule
Ideal HDR & 75.33 & 63.52 & 52.57 & 75.33 & 63.52 & 52.57  &  75.33 & 63.52 & 52.57  \\  
\bottomrule
\end{tabular}}
\label{tab:yolo} 
\end{table}

\begin{figure}[!b]
    \centering
    \includegraphics[width=\linewidth]{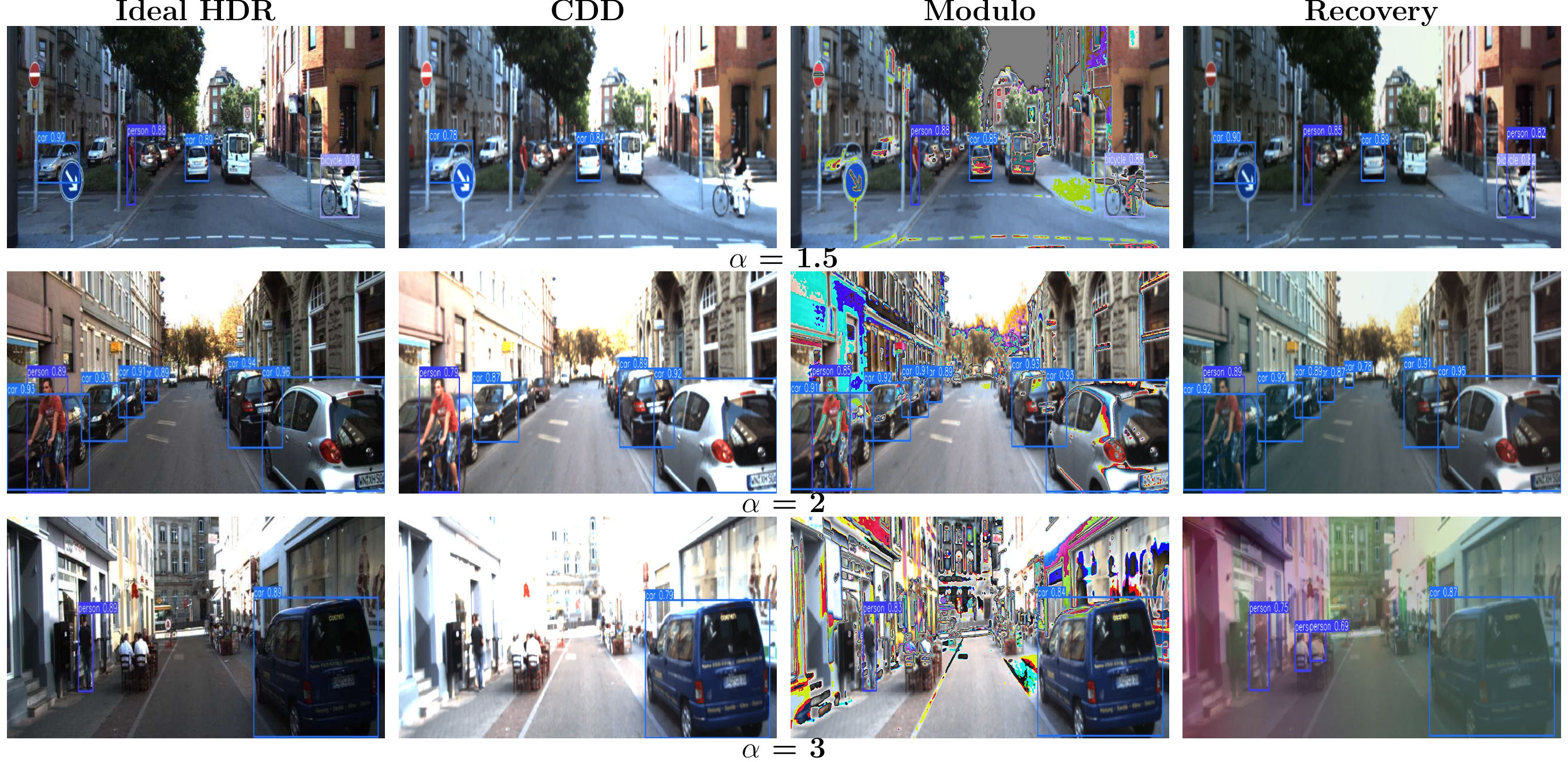} \vspace{-2em}
    \caption{Comparative of input modalities applied to a real urban scene with YOLO object detection. The first row shows images with saturation $\alpha = 1.5$, the second with $\alpha = 2$, and the third with $\alpha = 3$. Each column represents different methods: GT, CDD, Modulo, and Recovery, visually demonstrating how saturation influences the detections.}
    \label{fig:yolo_image} \vspace{-2em}
\end{figure}

\bacca{The numerical findings are presented in Table \ref{tab:yolo}. In particular, we notice that when $\alpha$ increases, the CCD results decrease drastically since more spatial zones are lost due to saturation. However, \textit{Modulo} and \textit{Recovery} provide better results and a lower drop in the metrics. Specifically, for small saturation levels $\alpha = 1.5$,  both \textit{Modulo} and \textit{Recovery} outperform the \textit{CCD} results where, for the \textit{Recovery}, provide similar results to ideal HDR. The visual results in Fig~ \ref{fig:yolo_image} further corroborate these quantitative results.
Interestingly, for the most challenging scenario $\alpha=3$, the proposed \textit{Recovery} method does not provide perfect recovery; however, it allows us to estimate better the object in the scene compared with CCD or raw modulo measurements. This general performance of the modulo sensor is based on addressing the saturation areas typically generated by excessive lighting from sources like car headlights or the sun. Information about these areas is often lost in the saturated image. In contrast, the modulo image captures these details, and our proposed method effectively recovers the wrapping levels, thereby preserving crucial information and improving object detection.}

\subsection{Discussion and Future Work}

We suggest that the proposed autoregressive phase unwrapping formulation within the USF framework, combined with incorporating pixel neighborhood vectorization for HDR modulo imaging, opens the way for innovative optimization-based unwrapping algorithms. These algorithms can use well-established signal priors in natural images, enhancing optimization strategies derived from the phase unwrapping problem. For instance, a promising research direction is the incorporation of denoising regularization such as that indicated for the first spatial finite difference~\cite{pineda2020spud, bacca2024deep}. This approach leverages powerful denoising algorithms and image-denoising neural networks to enhance HDR modulo imaging performance under noisy conditions without model retraining.

Furthermore, since our proposed method handles multi-channel and multi-frame HDR image restoration, the buffering of multi-frame images or burst images could further enhance HDR image quality for HDR video restoration, as presented in the literature~\cite{xiao2024deep, li2015multiframe, hasinoff2016}. Finally, an intriguing research path could involve the joint and iterative estimation of the HDR image and the stripe artifact map. This could be achieved by incorporating iterative image de-striping algorithms like those used in remote sensing images. Even more, one could explore different ordering paths in the image vectorization by neighborhood pixels. This alternative exploits the specific spatial structures presented in each image. Our experimental results indicate that no ordering path yields the best performance; it varies based on the image structure, wrapping location, and specific saturation factor.

\section{Conclusion}

This work introduces an autoregressive phase unwrapping method for HDR image reconstruction from modulo measurements. It employs high-level finite differences for 2D images, providing efficient solutions in the discrete cosine domain with a stride removal algorithm based on spatial sparsity. We show improved HDR restoration performance and evaluate our method against state-of-the-art optimization and deep learning HDR restoration algorithms using first-finite differences. We highlight the potential of modulo-ADC sensors and our HDR recovery algorithm in enhancing YOLO detector accuracy under overexposed conditions without retraining.


%
%
\bibliographystyle{splncs04}
\bibliography{main}

\end{document}